\documentclass[10pt,notitlepage]{iopart}
\usepackage{a4,epsfig}
\usepackage{url}
\usepackage[dvipsnames]{xcolor}
\usepackage{lineno}
\usepackage{multirow}
\usepackage{iopams}
\usepackage[toc,page,title]{appendix}
\usepackage[normalem]{ulem}
\begin{document}

\title{The MBTA Pipeline for Detecting Compact Binary Coalescences in the Third LIGO-Virgo Observing Run}
\author{
F~Aubin$^1$, F~Brighenti$^2$, R~Chierici$^3$, D~Estevez$^4$, G~Greco$^2$, G~M~Guidi$^2$, V~Juste$^4$,
F~Marion$^1$, B~Mours$^4$, E~Nitoglia$^3$,
O~Sauter$^1$\footnote{Present address: University of Florida Mechanical \& Aerospace Engineering, PO Box 116250, Gainesville FL, 32611, USA},
V~Sordini$^3$\footnote{Corresponding author}
}
\address{$^1$Laboratoire d'Annecy de Physique des Particules (LAPP), Univ. Grenoble Alpes, Universit\'e Savoie Mont
Blanc, CNRS/IN2P3, F-74941 Annecy, France}
\address{$^2$Universit\`a degli Studi di Urbino 'Carlo Bo,' I-61029 Urbino, Italy}
\address{$^3$Institut de Physique des 2 Infinis de Lyon (IP2I) - UMR 5822, Universit\'e de Lyon, Universit\'e Claude
 Bernard, CNRS, F-69622 Villeurbanne, France}
\address{$^4$Universit\'e de Strasbourg, CNRS, IPHC UMR 7178, F-67000 Strasbourg, France}

\begin{abstract}
We describe the Multi-Band Template Analysis (MBTA) search for gravitational waves signals from coalescences of compact objects in the LIGO-Virgo data, at the time of the third observing run (2019-2020), both for low-latency detections and for offline analysis. Details are given on the architecture and functioning of the pipeline, including transient noise mitigation strategies, parameter space for the searched signals, detection of candidates and evaluation of a false alarm rate associated to them. The performance of the low-latency search is demonstrated based on the LIGO-Virgo third observing run, during which MBTA has contributed to 42 alerts, submitting candidates with a median latency of 36 seconds. The performance of the offline search is illustrated on a subset of data collected during the second LIGO-Virgo observation run in 2017, and are quantified based on injections of simulated signal events on the same data.
\end{abstract}

\ioptwocol

\section{Introduction} \label{introduction}
The Multi-Band Template Analysis (MBTA) pipeline has been used to perform low-latency searches for compact binary coalescences (CBCs) ever since the late operation of the first generation of interferometric gravitational wave (GW) detectors (LIGO and Virgo science runs S6 and VSR2/VSR3)~\cite{LIGO_S6, Virgo_SR, S6_online, O1_BNS_NSBH, O2_online}. Its main features at the onset of the advanced detector era are described in~\cite{MBTA_2016}. The landscape of GW detection has dramatically changed since then, with the numerous CBC detections made with the advanced LIGO \cite{TheLIGOScientific:2014jea} and Virgo \cite{TheVirgo:2014hva} detectors, from a variety of sources involving black holes or/and neutron stars~\cite{GWTC1, GWTC2}. The searches performed by the LIGO-Virgo Collaboration include other pipelines used for online and/or offline modeled CBC GW searches, namely GstLAL~\cite{gstlal}, PyCBC~\cite{pycbc_live}, and SPIIR~\cite{spiir}, as well as the unmodeled search cWB \cite{cwb}.

Over the past years, we have improved the MBTA pipeline in various ways, driven both by the specific needs of the low-latency search designed to enable electromagnetic (EM) follow-up of GW candidates and by the desire to use the pipeline offline to produce future GW transient catalogs. While the follow-up goal initially lead to focusing mainly on sources potentially bright in the EM spectrum, i.e. containing at least one neutron star, the diversity of observed astrophysical sources~--~including heavy binary black hole (BBH) mergers~--~lead to exploring a parameter space extending to higher masses. To optimize the EM follow-up, we also improved the interface of the pipeline to the low-latency sky localization tool~\cite{Bayestar}, in order to provide more accurate sky maps. The online and offline goals met in requiring better noise rejection tools to improve detection efficiency, and in having to deal with a network of detectors with heterogeneous sensitivities. The paradigm used for noise rejection has shifted from applying thresholds on some discriminating observables to using those to weight the statistics used to rank candidates, in order to allow for a smoother transition from the signal-dominated regime to the noise-dominated regime.

MBTA performs a coincident analysis, so each detector in the network is analyzed separately before the results are combined to identify coincident events. It uses the matched filter technique, but splits it in two frequency bands to reduce the computational cost \cite{mbtamoriond2003}. The data selection and processing prior to analysis is described in section~\ref{data}. Section~\ref{templates} defines the parameter space covered by the search. Section~\ref{singletriggers} addresses the production and treatment of single-detector triggers, while section~\ref{coincidences} details how coincidences are extracted and section~\ref{far} how a false alarm rate is associated to them. Section~\ref{architecture} outlines the architecture of the pipeline in its online and offline configurations. Finally, section~\ref{performances} illustrates the performance of the pipeline in its low-latency as well as offline application.

\section{Data preprocessing} \label{data}
MBTA analyzes properly calibrated and conditioned time-domain strain data from times when any of the LIGO/Virgo detectors is in nominal
observing mode. These data are provided with data quality information \cite{LIGOScientific:2019hgc} in the form of veto flags indicating either severe problems affecting
the data (CAT1 vetoes) or likely problems revealed by auxiliary channels known to correlate to the strain channel (CAT2 vetoes). CAT2
vetoes are usually available with high latency and are used by MBTA only in the offline analysis.

Before processing the data through matched filtering, MBTA applies two preprocessing steps: decimation and gating. The former re-samples the
data at $4096~\mathrm{Hz}$ after suitable low-pass filtering in the frequency domain. The latter uses a specific procedure to remove short
stretches of bad quality data likely to harm the search. It is based on a similar idea and goal as the gating procedure developed by
PyCBC~\cite{pycbc_2016}, but uses a different implementation. The data are smoothly set to $0$ (using Tukey windows with a transition
time of $0.3125~\mathrm{s}$) whenever a CAT1 veto flag is active, or a CAT2 veto flag is active (in the case of the offline analysis), or a
snapshot estimate of the detector sensitivity crosses some adaptive threshold.

The instantaneous sensitivity is assessed from the reach to binary neutron star sources, the so-called BNS range, estimated here at
a relatively high sampling rate of $32~\mathrm{Hz}$. In order to reconcile fine sampling and good accuracy, the BNS range is computed from
an estimate of the current power spectral density (PSD) built, at each frequency, as the maximum between the result of a $0.25~\mathrm{s}$
fast Fourier transform (FFT) of the data and the median of all such FFTs over the past $10~\mathrm{s}$. The gating procedure is activated
whenever the range value gets below a threshold defined as $60\mathrm{\%}$ of the median range over the past $10~\mathrm{s}$. The minimal
amount of data affected by the gating is $\sim 0.66~\mathrm{s}$ (in the case of a single range sample below threshold), which is short
enough to be useful in the case of brief glitches overlapping astrophysical signals, like the loud transient noise affecting L1 data
$1.1~\mathrm{s}$ before the end of the GW170817 signal (see section~\ref{performances}). As part of the O3 online analysis, less than
$0.2\mathrm{\%}$ of science data have been gated, with a typical dead time shorter than $2~\mathrm{s}$.

This method is potentially unsafe since it is based on a local estimate of the BNS range, and gating could be triggered by loud astrophysical
signals. Tests with simulated signals have shown that strong and short signals from high-mass, nearby BBH sources are indeed liable to be
missed due to the gating. To mitigate this issue, the relevant part of the BBH parameter space is also analyzed without applying the
gating procedure, although with higher signal-to-noise thresholds (see sections~\ref{templates} and~\ref{singletriggers}).

Matched filtering requires an accurate estimate of the detector PSD. MBTA uses PSDs estimated from preprocessed data (i.e. after
decimation and gating). The estimate is based on successive, partially overlapping FFTs. The FFT size is configuration dependent and
ranges from seconds to hundreds of seconds. To achieve suitable accuracy, we use the median value of the FFTs (correcting for the inherent bias~\cite{Allen:2005fk}) over a time which is
also variable and ranges from $1000$ to $4000~\mathrm{s}$.

\section{Parameter space and template banks} \label{templates}
\begin{table*}
\caption{Division of the parameter space into regions and parameters used to generate the corresponding template banks.}
\label{table_banks}
\centering
\begin{tabular}{|c|c|c|c|c|c|c|c|c|}
\hline
Region & $m_1/\mathrm{M}_{\odot}$ & $m_2/\mathrm{M}_{\odot}$ & $\left | \chi_{1,z} \right |_{\mathrm{max}}$ & $\left| \chi_{2,z} \right|_{\mathrm{max}}$ &
 $f_0$ (Hz) & $f_c$ (Hz) & Waveform & Waveform \\
 & & & & & & & (bank) & (analysis) \\ \hline
1 & [1;2] & [1;2] & 0.05 & 0.05 & 25 & 80 & TaylorF2 & SpinTaylorT4 \\ \hline
2 & [1;2] & [2:100] & 0.05 & 0.997 & 23 & 85 & SEOBNRv4\_ROM & SEOBNRv4 \\ \hline
3 & [2;195] & [2:195] & 0.997 & 0.997 & 23 & 85 & SEOBNRv4\_ROM & SEOBNRv4 \\
&\multicolumn{2}{|c|}{$(m_1+m_2)< 200\,\mathrm{M}_{\odot}$} & & & & & & \\ \hline
\end{tabular}
\end{table*}

MBTA relies on models of CBC signals to generate simulated waveforms, called {\em templates}, used to match filter the data. A bank of such templates is needed to cover the parameter space explored by the search. In O3, as in O1 and O2, MBTA considered a 4-dimensional space, consisting of the masses of the compact objects $m_1$, $m_2$ and their dimensionless spins assumed to be parallel to the orbital angular momentum $\chi_{1,z}$, $\chi_{2,z}$. It has been shown that taking spin misalignment into account does not result in much higher detection efficiency \cite{Abbott_2008, Van_Den_Broeck_2009, Dal_Canton_2015, Abbott_2016}. Templates are distributed in such a way to ensure that the loss in signal-to-noise ratio (SNR) from using a discrete sampling of the parameter space is limited (typically less than 3\%) for any astrophysical signal.

In O3, MBTA covers an extended mass space with respect to O1 and O2: single (detector frame) masses range from 1 to 195~$\mathrm{M}_{\odot}$ with total mass less than 200~$\mathrm{M}_{\odot}$ (compared to 100~$\mathrm{M}_{\odot}$ in O2). The expansion to higher masses was motivated by the observation of high mass sources in O2~\cite{GWTC1} and early O3~\cite{GW190521} as well as the more sensitive detectors reaching sources at higher redshift in O3, translating into higher detector frame masses. A broad tiling is appropriate for the high-mass part of the parameter space, which means that the search sensitivity does not drop sharply beyond the high mass cutoff but somewhat extends to heavier sources.

Regarding spins, $\chi_{1,z}$ and $\chi_{2,z}$ cover ranges matching current astrophysical expectations
for neutron stars in binaries ($\left | \chi_{z} \right |<0.05$) and black holes ($\left | \chi_{z} \right | \lesssim 1$).
We consider that a compact object with mass above 2~$\mathrm{M}_{\odot}$ can be a black hole, therefore allowing for the broad spin range,
and we use the narrow spin range only for objects below 2~$\mathrm{M}_{\odot}$.

The above considerations lead us to split the mass parameter space into primarily three regions and generate region-specific template banks, as summarized in table~\ref{table_banks}. The regions can be seen as matching the BNS, NSBH and BBH spaces (and we will occasionally refer to them as such), with the caveat that the transition from NS to BH occurs at 2~$\mathrm{M}_{\odot}$ instead of the more typical 3~$\mathrm{M}_{\odot}$.
In addition, we duplicate a small fraction
of the region 3 bank to make up the so-called region 4, with templates having a short duration (typically less than $\
\sim 3~\mathrm{s}$ from 21~Hz). Region 4 is used to analyze data without the gating procedure (see sections~\ref{data} and~\ref{singletriggers}).

Signal waveforms of binary coalescences are generated
by treating the inspiral phase at a given post-Newtonian (PN) order, and in some cases by including the merger and ringdown derived from numerical relativity.
For the simulation of these sources we use the SpinTaylorT4 approximant \cite{Buonanno, Buonanno_2003} for BNS as implemented in the
LALSimulation library \cite{LALSimulation}, whereas we use SEOBNR \cite{Bohe_2013, Bohe_2017} elsewhere.
We employ orbital PN phase corrections up to order 3.5 and aligned spin-orbit and spin-spin terms up to 3.5.  For the BNS waveforms only the inspiral phase is generated, since the analysis is not expected to be sensitive to the high
frequency ringdown features.

As the MBTA analysis is split over two frequency bands (see~\cite{MBTA_2016} for more details), it needs both a bank of templates covering the full frequency band starting at frequency $f_0$, called virtual templates (VT), and banks of real templates (RT)~---~actually used for matched filtering~---~for the low-frequency band (from $f_0$ to the crossing frequency $f_c$) and the high-frequency band (above $f_c$). Different regions use different values of $f_0$ and $f_c$ to account for the dependence of the SNR distribution on the mass of the system.

The production of the VT and RT banks makes use of a stochastic generation method~\cite{Babak, Privitera}. Different waveform models are used for region 1 and regions 2 and 3, as in the latter case the merger and ringdown parts of the waveform need to be taken into account. The TaylorF2 model~\cite{Buonanno} is used for BNS and the SEOBNRv4 ROM model~\cite{Bohe_2013, Bohe_2017, Purrer} for NSBH and BBH. The actual analysis uses similar models operating in the time domain. Each VT is associated to a low-frequency RT and a high-frequency RT, identifying the RTs that best match the VT.

Table~\ref{table_offline} gives the number of templates used in each region for the O3 offline analysis. The O3 online analysis used similar numbers, with two minor differences: a pre-O3 reference sensitivity curve was used, resulting in smaller template banks, and slightly different configurations, especially regarding values for the maximum total mass in region 3, $f_0$, $f_c$ and the fraction of region 3 included in region 4.

\begin{table}
\caption{Number of templates used in the O3 offline analysis. $f_{low}$ is the low frequency cutoff actually used in the analysis,
which in the BBH case is slightly lower than the value of $f_0$ used to generate the banks. \label{table_offline}}
\centering
\begin{tabular}{|c|c|c|c|c|}
\hline
Region & VT & RT LF & RT HF & $f_{low} (\mathrm{Hz})$ \\\hline
1 & 27441 & 7806 & 1901 & 25 \\\hline
2 & 524016 & 133659 & 12200 & 23 \\\hline
3 & 176535 & 50429 & 6296 & 21 \\\hline
4 & 9000 & 2923 & 430 & 21\\\hline
\end{tabular}
\end{table}

\section{Single detector triggers} \label{singletriggers}

For each detector and each template in the virtual template bank, MBTA computes the outcome of the matched
filtering process applied to the data of that detector as a function of time. The processing is performed via
FFTs, using in-phase and in-quadrature instances of the real templates in
the two frequency bands. It is followed by a combination step that coherently sums the results from
the low-frequency and high-frequency bands, by applying the proper translation in time and rotation in
phase, and provides an SNR as a function of time.
For each unit stretch of processed data (covering an amount of time defined by the configuration-dependent sizes of the FFTs), the maximum SNR $\rho$ is checked against the single-detector ranking statistic threshold $\rho_{\rm min}$ and a trigger is recorded if $\rho \geq \rho_{\rm min}$.
A basic acceptance criterion is applied at this stage, using a loose cut on the value of a 2 degrees-of-freedom
$\chi^2$ observable designed to compare the distribution of SNR between the two frequency bands to the expectation
from an astrophysical signal~\cite{chi2_Allen}. For more details on the SNR computation and the $\chi^2$ cut,
see~\cite{MBTA_2016}.

During O1 and O2, MBTA used another consistency test based on the evolution of the SNR around the time of the
maximum $\rho$, applying a cut on an observable comparing the behavior close to the maximum and further away
from it.
A more effective approach consists in weighting the SNR according to the value of a discriminating observable, following an idea proposed in~\cite{S3_S4_CBC}, as this allows a better handling of the regime where signal and noise overlap.
For O3 we adopt this approach, using an observable which still makes use of the SNR time series but is a variant of the autocorrelation-based least-squares test described in~\cite{gstlal}. We define:
\begin{equation}
\rho_{rw} = \left\{ \begin{array}{ll}
 \rho &\mbox{ if $\xi^2_{PQ} \leq 1$} \\
  \rho \ \left(\frac{A +  \left. \xi^2_{PQ}\right.^{\alpha}}{A+1}\right)^{-1/\beta} &\mbox{ if $\xi^2_{PQ} > 1$}
       \end{array} \right.
\end{equation}
with $\xi^2_{PQ}$ a discretized version of:
\begin{equation}\setlength\arraycolsep{0.01em}
\xi^2_{PQ} = \frac{1}{2 \Delta t} \int_{t_0-\frac{\Delta t}{2}}^{t_0+\frac{\Delta t}{2}} \left\|
\left(\begin{array}{c}\rho_P(t) \\ \rho_Q(t)\end{array}\right) -\rho
\mathcal{R}
\left(\begin{array}{c}A_P(t-t_0) \\ A_Q(t-t_0)\end{array}\right)
\right\|^2 dt
\end{equation}
where $\rho$ and $\phi$ are the modulus and phase of the complex SNR of the trigger recorded at time $t_0$, $\mathcal{R}$ is the rotation matrix associated to $\phi$,
$\rho_P$ and $\rho_Q$ are the matched filter outputs with the in-phase and in-quadrature templates respectively,
$A_P$ and $A_Q$ are the autocorrelations of the template with its in-phase and in-quadrature avatars,
with normalized maximum at $t=t_0$. Parameters are chosen as $A=10$, $\alpha=5$, $\beta=8$.

In order to identify times when the detectors show evidence of poor data quality, we compute a new observable
sensitive to an excess in the rate (in hertz) of triggers $R_{(\rho \geq \rho_{\min})}$, compared to the rate of
surviving triggers once the $\chi^2$ cut is applied and the reweighted SNR is considered
$R_{(\rho_{rw} \geq \rho_{\min})}$:
\begin{equation}
E_R(t_0) = \mathrm{median}_{[t_0 + t_{\mathrm{offset}} - 10s, t_0 + t_{\mathrm{offset}}]}\ \delta R(t)
\end{equation}
with
\begin{equation}
\delta R(t) = \frac{R_{(\rho \geq \rho_{\min})}(t) - R_{(\rho_{rw} \geq \rho_{\min})}(t)}{R_{(\rho \geq \rho_{\min})}(t)}
\end{equation}
$\delta R$ is computed at 8~Hz and $E_R$ at 1~Hz.
In the same spirit as the reweighted SNR, a new ranking statistic is defined that takes into account $E_R$,
in order to penalize triggers occurring at the time when there is an excess of poor quality triggers:
\begin{equation}
\rho_{rw,E_R} = \left\{ \begin{array}{ll}
 \rho_{rw} &\mbox{ if $E_R \leq 0.3$} \\
  \rho_{rw} \ [1 - A (E_R - 0.3)^{\alpha}] &\mbox{ if $E_R > 0.3$}
       \end{array} \right.
\end{equation}
With the current definition of $\rho_{rw}$ and upstream cuts, we find that the following values for the parameters
are suitable: $A=1$, $\alpha=2$, $t_{\mathrm{offset}}=0$ online (to minimize latency) and
$t_{\mathrm{offset}}=7\ \mathrm{s}$ offline.

\section{Coincidences} \label{coincidences}
Candidate events are defined as time coincidences between single-detector triggers found with the same template. Only triggers above $\rho_{\min}=$ 4.8 (4.5) are considered for the BBH and NSBH (BNS) search.

The first step is to look for double coincidences, considering each pair of detectors to produce “HL”, “HV” and “LV” events.
A tight cut of $w_{\mathrm{HL}} = \pm$ 15~ms is applied on the time of flight for HL coincidences and a looser cut of $w_{\mathrm{HV}} = w_{\mathrm{LV}} = \pm$ 35~ms for HV and LV.
The second step is to look for triple coincidences:
this is implemented by searching for HL and HV events that share the same H single-detector trigger.

In order to improve sky localization~\cite{Fujii,EM_user_guide}, so called pseudo-triple events are also made from double events
observed during triple detector times, by searching for a same-template, sub-threshold signal in the third detector.
Parameters like the combined SNR or sky maps of these pseudo-triple events are made using the three-detector information,
while their significance is that of the double event.

Finally, a clustering across the template bank is performed to group coincidences that are
close in time (typically within 20~ms of any trigger in the cluster), and therefore likely to originate from the same event.
The candidate event is then characterized only by considering the coincidence with the highest ranking statistics.

Triggers coming from astrophysical events are expected to show correlations across detectors, not just between their arrival times but also regarding the phase and amplitude of the signal.
This information is used when building the combined ranking statistics of double coincidences by adding to the quadratic sum of the individual ranking statistics a term including probabilities quantifying how likely the measured parameters are for a population of sources~\cite{pycbc_2017}:

\begin{equation} \label{eq:cRS}
\rho_{RS,ij}^2 = \rho_{rw,E_R,i}^2 + \rho_{rw,E_R,j}^2 + 2 \ln(P_{\Delta t_{ij}} P_{\Delta \phi_{ij}} P_{RA_{ij}})
\end{equation}
\begin{equation} \label{eq:cRST}
\rho_{RS,ijk}^2 = \rho_{RS,ij}^2 + \rho_{RS,ik}^2 - \rho_{rw,E_R,i}^2
\end{equation}

The probabilities $P_{\Delta t_{ij}}$, $P_{\Delta \phi_{ij}}$ and $P_{RA_{ij}}$ are driven by the location, orientation and sensitivity of the detectors (labeled as $i$, $j$, $k$). They are built for the time difference, phase difference and relative amplitude of detectors $i$ and $j$, using a population of sources uniformly distributed in volume and neglecting any correlation between these parameters. In order to account for the detector resolution, a 0.5~ms jitter is applied on the time of flight, 0.15 on the amplitude ratio and 0.5~rad on the phase difference.
A threshold, with a typical value of 7, is finally applied on the combined ranking statistics $\rho_{RS,ij}$.

\section{Computing the false alarm rate} \label{far}

Coincidences are assessed by determining their false alarm rate (FAR),
i.e. the rate of coincidences at least as significant produced by the search from noise triggers.

There are two main steps to compute this quantity, detailed in this section.
The first one is to determine an initial FAR, independently for each single search (defined as a given
region and a given type of double or triple coincidence).
This is done by building a background model using random coincidences of single-detector triggers.
The second step is to account for these multiple searches running in parallel, to provide the final
FAR relative to the overall coincidence time,
defined as the time with at least two operating detectors.
This is done by applying trial factors on the previously computed FAR.

\subsection{FAR for a single search}
Within a single search, the FAR of a coincidence is defined as the rate of coincidences from noise triggers
with ranking statistics values as high or higher than that of the considered coincidence.
Its computation is done by assuming that the detectors are independent.
It uses single-detector triggers, which are largely dominated by noise given the current astrophysical rate of events.
Time windows a few seconds wide that cover potential real events~---~
identified as true coincidences with a
combined ranking statistics larger than 10~---~are excluded from this sample.
In order to keep the background computation manageable, a down-sampling of the low SNR events, with proper reweighting, is performed
(the number of events used per unity of SNR is roughly constant below 6.4).

A sample of single-detector triggers is then built using the last 24 hours of single-detector up-time for the online operation,
or about six days for the offline runs.
Starting from this sample, fake coincidences are built by making all possible combinations of single-detector triggers with identical template from different detectors, independently of their arrival times.
Any significant detector non-stationarity, or possible correlation, is averaged out in this process.

To get the combined ranking statistics of a fake coincidence,
the parameter consistency checks are applied using the single trigger parameters for the phase and amplitude,
and a random value within the allowed range for the time of flight.
The FAR at a given combined ranking statistics threshold $FAR_{ij}(\rho_{RS,ij})$ is then computed by counting the number of fake coincidences above that threshold $N_{ij}(\rho_{RS,ij})$,
and renormalizing it by the coincidence time window $w_{ij}$,
divided by the product of the analyzed times for the two detectors in the pair $ij$:

\begin{equation} \label{eq:farHL}
FAR_{ij}(\rho_{RS,ij}) = N_{ij}(\rho_{RS,ij}) w_{ij}/(T_i T_j)
\end{equation}

The FAR for triple events can be derived from the FAR of its HL and HV events,
correcting for the probability to get an H trigger since it is entering both in $FAR_{HL}$ and $FAR_{HV}$.
More specifically (see simplified equation~\ref{eq:farHLV} which ignores the $RS$ subscripts),
$FAR_{HLV}$ is built by integrating the product of the FAR distributions for HL ($ij$) and HV ($ik$) coincidences,
renormalized by the triple detector time ($T_{ijk}$) and by the distribution of the single H ($i$) events taken at a
ranking statistic derived from equation \ref{eq:cRST}:
$\rho_{i}^2 = \rho_{ij}^2 + \rho_{ik}^2 - \rho_{ijk}^2$.

\begin{equation} \label{eq:farHLV}
FAR_{ijk}(\rho_{ijk}^2) =
\int \int_{\rho_{ij}^2, \rho_{ik}^2}
\frac{FAR_{ij}({\rho_{ij}^2}) FAR_{ik}({\rho_{ik}^2})}
{N_i (\rho_{ij}^2 + \rho_{ik}^2 - \rho_{ijk}^2 )/T_{ijk}}
\end{equation}

The clustering procedure then reduces the number of candidates, making each of them more significant.
For this reason, the FAR obtained before clustering is scaled by a factor $\kappa_{\rm cluster}$ equal to the average number of clustered events divided by the number of events before clustering.
This factor was estimated to be $0.59$ for BNS and $0.44$ for BBH and NSBH for the online configuration.
For the offline running, the slight dependence with the event combined ranking statistic is taken into account assuming a linear behavior. In the signal-dominated region the values are compatible with the correction used online.

\subsection{FAR for the overall search}

Without considering region 4, the overall pipeline is composed of 12 individual searches (3 regions times up to 4 coincidence types). After the first step, each single search brings the same background distribution, in terms of the
rate of noise coincidences with FAR below a given threshold. The most straightforward approach would be to combine
them by simply rescaling all FAR with a trial factor accounting for the various single searches. While we take this
simple approach for the regions, we take a different path for the coincidence types, including some rough
astrophysical priors to optimize the overall search, as explained in the following.

Within a region, HL, HV, LV and HLV events are not always possible, since a detector could be down independently of the status of the other detectors.
The various coincidence types do not have the same
ability to capture astrophysical signals, therefore we allocate to each of them a ``background budget'' related
to that ability.
To account for this effect, the value of the FAR is divided by trial factors $\kappa_{\rm coinc}$ that capture the relative searched volume of each type of coincidence.
They are estimated using an astrophysical source population simulation and counting the number of sources expected to be detected as different coincidence types.
During triple detector times, the typical values used for $\kappa_{\rm coinc}$ are 0.909, 0.002, 0.007 and 0.083 for the HL, HV, LV and HLV exclusive types of coincidence.
For double detector time, during the online operation of O3 the trial factor of the only possible pair was set to 1, even for the time with only one LIGO detector, time less likely to detect real events than triple detector time.
This was improved offline by using a reduced trial factor equal to the fraction of volume probed by the search of the two active detectors relative to the three-detector case.  For instance, during HV only time, $\kappa_{\rm coinc}=0.002+0.083$.
An overall scaling factor depending on the fraction and type of two-detector time is then applied during all times to produce a FAR normalized to the overall coincidence time.

On top of the detector combination, we apply a $\kappa_{\rm region}=1/3$ trial factor to account for the parallel searches running on the three regions. The fourth region that is looking for BBH events on ungated channels is ignored since it produces only very significant events, given its high threshold.

Overall, the inverse false alarm rate (IFAR) of a cluster is computed from the FAR for that particular type
of coincidence in that particular region according to:
\begin{equation}
\mathrm{IFAR} = \frac {\kappa_{\rm region}\ \kappa_{\rm coinc}}{ \kappa_{\rm cluster}\ \mathrm{FAR}(\rho_{RS})}
\end{equation}
IFAR cumulative distributions such as the ones presented in figures~\ref{fig:IFARCumulO3A} and ~\ref{fig:IFARCumulO2_offline} in section \ref{performances} are used to assess the consistency of the FAR computation and to illustrate our ability to understand the background.
At a given IFAR threshold, the red dots show the number of candidate events with an IFAR equal or larger than the threshold.
The grey bands represent the 1$\sigma$, 2$\sigma$ and 3$\sigma$ statistical Poissonian uncertainties around the expected curve that is normalized to the total coincidence time of the analyzed period.
True astrophysical events are expected to show up as outliers in this plot, at high values of IFAR.

\section{Pipeline architecture} \label{architecture}
The pipeline starts from the $h(t)$ streams coming from the three detectors.
The first step is to apply, on each data stream, the preprocessing (down-sampling and gating) described in section~\ref{data}.

Then, with version 3 of the pipeline used during the first half of O3,
the matched filtering was performed separately on each data stream,
by multiple processes running in parallel.
Each single filtering process passed the triggers exceeding a ranking statistic threshold
to a coincidence process building double and triple coincidences.
That process also performed the FAR estimation using the received single-detector triggers.
The coincidences were then clustered together by another set of processes,
and results passed to the process in charge of the connection to GraceDB \cite{gracedb} and recorded.
Version 3 of the pipeline had the limitation that the information provided for sky localization
was restricted to the signals observed in coincidence above the selection threshold, missing
sub-threshold information from the third detector, if active at the time.

With version 4 of the pipeline, put into production at the beginning of the second half of O3 (November 2019),
the coincidence step is moved upstream in the processes in charge of the matched filtering, which now process
the two frequency bands of the three detectors up-front.
Then, if a single trigger is recorded in a detector, the two frequency bands for the other available detectors are recombined in order to provide the sub-threshold information for the triggering template.
Those processes now also take care of the coincidence step and monitor the single-detector triggers, which are
used by another process determining on a periodic basis the relation between FAR and combined ranking statistic,
as described in section~\ref{far}.
Finally, a last process assigns a FAR to each coincidence and performs clustering.

The MBTA pipeline is implemented as a set of applications developed in C, sharing libraries with the Virgo online system.
This allows a smooth connection to the Virgo real time system, as well as the data coming from LIGO.
Communication between pipeline elements relies on TCP/IP connections for distant processes,
shared memories for processes running on the same machine, or simple files for offline processing.
The connections, as well as the parameters of the processes,
like data channel names or threshold values, are defined via configuration files allowing
to use the same binary code either online or offline.
The pipeline uses the frame format and associated software \cite{frlib} to save both data streams and candidate
events.
The input is either live data coming for the three detectors or reprocessed $h(t)$ data read from disk.  Details about the online and offline implementation are given in \ref{app:implementation}.

\section{Performance} \label{performances}
This section illustrates the performance of the MBTA analysis, first for the online running, based on the first half of the O3 observation campaign, and then for the offline running, on a partial data set from the 2017 O2 observation run.
\subsection{Online search}

\begin{figure} [h]
\centerline{\includegraphics[width=0.5\textwidth]{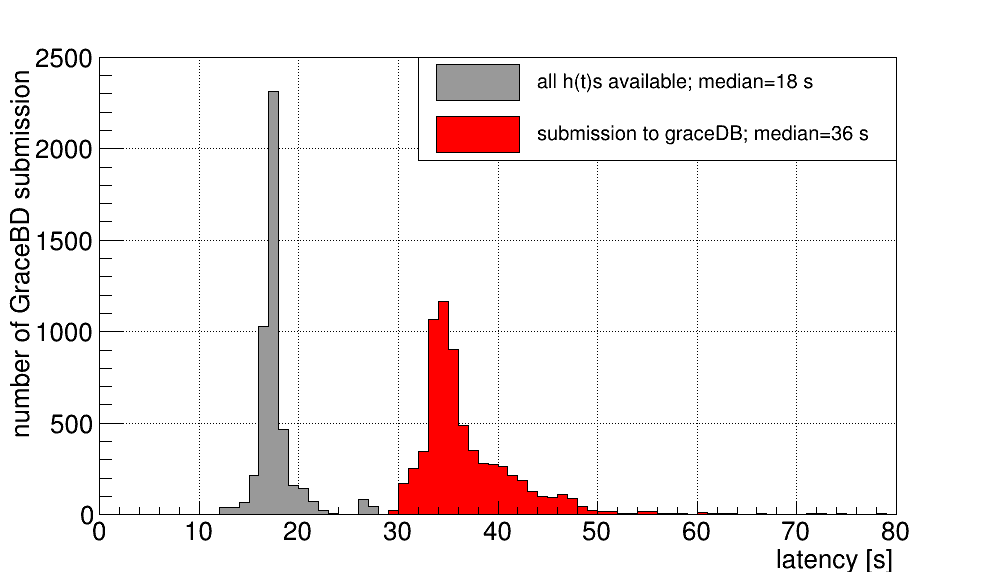}}
\caption{Latency for the h(t) data readiness (grey), i.e. the start of MBTAOnline, and for the candidates submission (red), i.e. the end of the MBTAOnline processing during O3.\label{fig:latency}}
\end{figure}

The online pipeline was fed by 1 second long frames arriving at the Virgo site with a latency of about 18 seconds,
see figure \ref{fig:latency},
driven by data collection and $h(t)$ reconstruction at each detector site, then transfer to the Virgo site.
The median value of the overall MBTA latency was 36 seconds,
with half of this time to get the $h(t)$ data.
The interaction with GraceDB, to complete the full upload of auxiliary files,
was taking a few to several more seconds, depending on the number of candidates submitted by the other pipelines.

\begin{figure} [h]
\centerline{\includegraphics[width=0.5\textwidth]{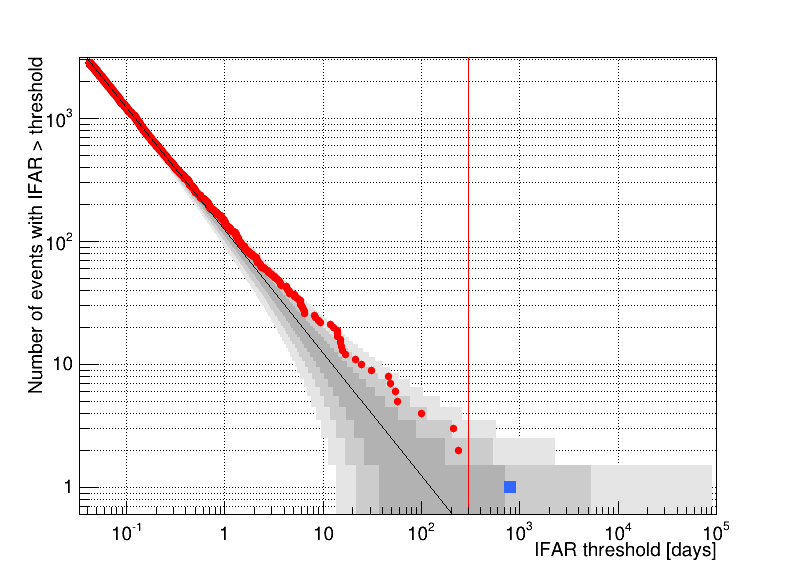}}
\caption{IFAR cumulative distribution of the MBTAOnline search over the period from May $2^{nd}$ 2019 to October $1^{st}$ 2019 with 24 O3a events of GWTC-2 removed, 16 of them above the IFAR threshold. The starting date of the period differs from the first day of O3a (April $1^{st}$ 2019) since May $2^{nd}$ is the time when MBTAOnline alerts were submitted for the whole parameter space. The red dots are all the coincidences produced by this pipeline without the events from GWTC-2 and the blue square is a retraction. The red line indicates the alert threshold used for O3 ($\mathrm{IFAR} = 10$ months), and the black line and gray bands show the expected behavior of stationary noise for the average number of events and $1\sigma$, $2\sigma$ and $3\sigma$ statistical Poissonian uncertainties.}
\label{fig:IFARCumulO3A}
\end{figure}

During O3, the MBTAOnline analysis contributed triggers to 42 low-latency alerts sent out by the LIGO-Virgo Collaboration. Among those triggers, 34 had a FAR below the alert threshold and 5 led to alerts being promptly retracted because of data quality issues indicating a likely terrestrial origin. The non-retracted MBTA candidates of O3a with FAR below alert threshold were confirmed by offline searches, which means they may have astrophysical origins \cite{GWTC2}. Figure~\ref{fig:IFARCumulO3A} shows that the analysis has produced in O3a a distribution of triggers consistent with expectations from stationary noise. Although compatible, one can notice that as the IFAR increases the distribution tends to deviate from the expectation value. This excess can be attributed to some variations of the detectors noise and perhaps to a few unidentified events.

The MBTA search does not yet assign a significance to single-detector triggers therefore during O3 submitted only coincidences to GraceDB. Nevertheless, for most of the public single-detector alerts (uploaded by GstLAL), MBTA detected a related trigger with similar SNR.

\subsection{Offline search}

The performance of the offline analysis is illustrated by applying it to a subset of O2 data, and is quantified based on simulated signals injected into the same O2 data set.
GW signals from compact binary coalescences are generated, for the three regions, in a random scan of masses, spins and distances in the interesting range (see table \ref{tab:InjPars}), and uniformly in inclination and position in the sky.

\begin{table} [h!]
\caption{Total mass ($m_{\rm tot}$) range, waveform generator, and distance ($D$) range
for the injections. The BBH and NSBH coalescences are simulated in two separate injection sets. Each sample is generated uniform in distance and total mass, and a weight based on the generated distance is applied to each event, to simulate an ensemble of injections uniform in volume.\label{tab:InjPars}}
\centering
\begin{tabular}{|c|c|c|c|}
\hline
Region & Waveform & $m_{\rm tot}$ ($\mathrm{M}_{\odot}$)& $D$ (Gpc)\\\hline
BNS & SpinTaylorT4 & 2 - 4 & 0.02-0.4 \\\hline
\multirow{2}{*}{NSBH} & \multirow{2}{*}{SEOBNRv4} & $<17$ & 0.07-0.5\\\cline{3-4}
 &  & $17-100$ & 0.07-1\\\hline
\multirow{2}{*}{BBH}& \multirow{2}{*}{SEOBNRv4} & $<30$ & 0.15-2\\\cline{3-4}
 & & $30-200$ & 0.15-6\\\hline
\end{tabular}
\end{table}

The randomly generated waveforms are superimposed to the data, considering only periods of good data-taking, at intervals of 12 s, amounting to a statistics of about 60000 injections per week. The possibility of long waveforms overlapping previous injections was tested to give a negligible impact in terms of the evaluation of the detection efficiency.
The presence of rare astrophysical signals in the data stream will have a negligible impact in the determination of the analysis performance.

For the purpose of quantifying the performance of the O3 offline analysis, MBTA has been used to analyze the O2 data collected from August 13, 2017 06:09:42 to August 24, 2017 02:41:22 (GPS time range 1186639800 - 1187577700), as available in the Gravitational Wave Open Science Center \cite{Abbott:2019ebz}.
The ability of the analysis to detect the injections is studied by looking for the candidate events given by MBTA in a 150~ms window around the injection.
The probability of detecting a signal from a source in a detector depends on their actual distance and their relative
orientation, i.e on their effective distance, defined as in \cite{thorne.k:1987}. For a coincidence this probability depends on the decisive distance: the second smallest effective distance, within the considered detectors.
The detection efficiency, defined as the fraction of simulated injections recovered by the analysis, goes to almost 100\%
for the injections with smallest decisive distances and reaches 50\% for sources at a decisive distance of 260 Mpc for
BNS, 450 Mpc for NSBH, and 1 Gpc for BBH.
When looking at the efficiency as a function of the actual distance to the source, regardless of its direction and orientation,
geometrical effects reduce the efficiency, which reaches 50\% for distances of 100 Mpc for BNS, 160 Mpc for NSBH, and 400 Mpc for BBH.

\begin{figure}[h!]
\centerline{\includegraphics[width=0.48\textwidth]{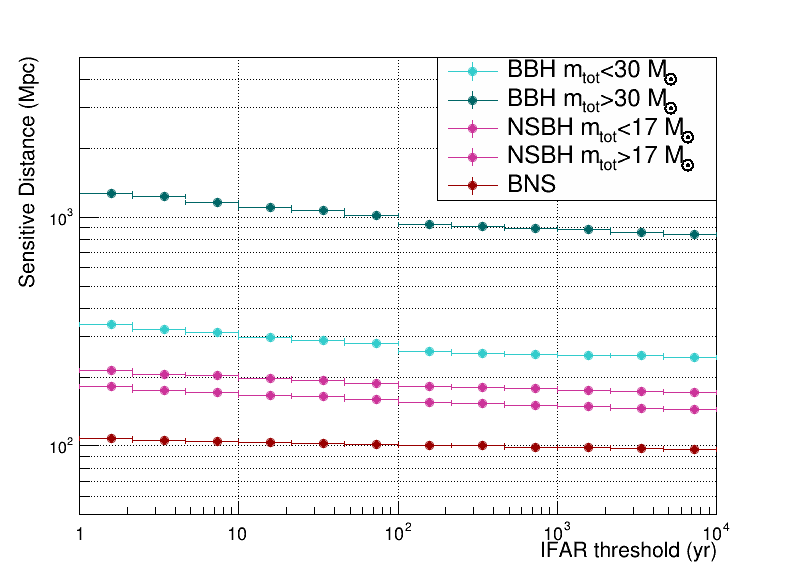}}
\caption{Sensitive distance of the MBTA offline search on a subset of O2 data, as a function of the selection cut on the inverse false alarm rate, for simulated GW signals from BNS, NSBH and BBH coalescences. For BBH and NSBH, the sensitive distance is shown separately for low and high total masses of the system.
\label{fig:SensDist_IFAR_O2_offline}}
\end{figure}

A way to quantify the distance at which the pipeline detects astrophysical signals is to determine the so-called sensitive distance: this is defined as the radius of a sphere whose volume is $\epsilon V$, with $\epsilon$ the selection efficiency (for a given IFAR threshold) and $V$ the volume in which injections are done, neglecting cosmological effects. This quantity
is shown in figure~\ref{fig:SensDist_IFAR_O2_offline}, as a function of the value of the IFAR threshold, for the three regions. Although the value of the sensitive distance depends on the simulations used to assess it, figure~\ref{fig:SensDist_IFAR_O2_offline} is a way to visualise the performance of the analysis.
In particular, the sensitive distance for NSBH signals is evaluated for a NS mass of 1.4~M$_{\odot}$ and a BH mass of 5, 10 and 30~M$_{\odot}$ and found to be about 160, 200 and 230 Mpc, respectively. Assuming similar sensitivity over the full O1 and O2, and no NSBH detection in those runs, these values are in reasonable agreement with the rate upper limits obtained by other pipelines and presented in figure 14 of \cite{GWTC1} for non-precessing sources.

\begin{figure}[h!]
\centerline{\includegraphics[width=0.48\textwidth]{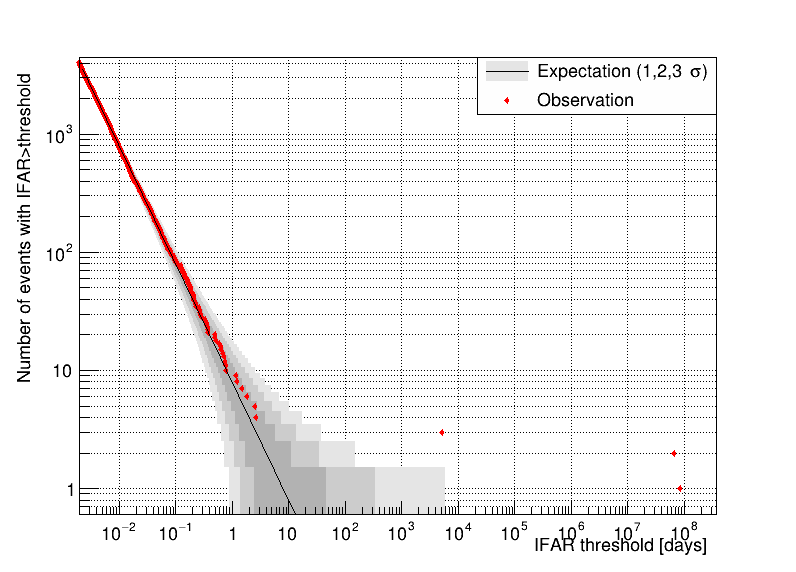}}
\caption{Inverse false alarm rate (IFAR) cumulative distribution for the MBTA offline running on the O2 data from August 13, 2017 06:09:42 to August 24, 2017 02:41:22. For a given IFAR threshold the red dots show the number of candidate events with IFAR equal or larger than the considered value. The black curve shows the expected number of events from background, and the grey bands represent the $1\sigma$, $2\sigma$, and $3\sigma$ statistical uncertainties around it (Poissonian errors are taken into account). The expectation is normalized to 8.05 days, the time during which at least two detectors where taking data of good quality.
\label{fig:IFARCumulO2_offline}}
\end{figure}

Once fully characterized on simulation, the search is run on the data to look for real astrophysical candidates. The cumulative distribution of the inverse false alarm rate for the considered observing period, corresponding to 8.05 effective days with at least two detectors functioning, is shown in figure~\ref{fig:IFARCumulO2_offline}. Most of the observations are in good agreement with the expectation from background, and three candidates stem out significantly. These events all have an inverse false alarm rate higher than 30 days, which was the threshold used to consider a detection in GWTC-1.
The details about such events are given in table~\ref{tab:O2offlineEvts}, which also contains the information about the confirmed detection to which they correspond \cite{LIGOScientific:2018mvr}. One additional event was observed during this period by the LIGO and Virgo collaborations, GW170818, with an SNR in Virgo of 4.2, a Hanford SNR of 4.1, and a Livingston SNR of 9.7. In the MBTA search we observe a corresponding single-detector trigger in Livingston with a SNR of 9.2. The SNR in the other two detectors are not high enough to build a coincidence.

\begin{table*} [h!]
\caption{MBTA offline search results for the period between August 13, 2017 06:09:42 and August 24, 2017 02:41:22. Three events are detected as coincidences with an inverse false alarm rate higher than 30 days, for those the table reports the date, UTC time, inverse false alarm rate, network SNR, chirp mass and corresponding identification of confirmed GW event.  The table also mentions the GW170818 detection from \cite{LIGOScientific:2018mvr}, for which MBTA detects a single trigger in Livingston. \label{tab:O2offlineEvts}}
\centering
\begin{tabular}{|c|c|c|c|c|c|}
\hline
Date & UTC time & IFAR (yr) & Network SNR & $\mathcal{M}^{det} (\mathrm{M}_{\odot})$ &GW Event \\\hline
Aug 14 2017 & 10:30:43 & $>2.30$ $10^{5}$ & 16.7 & 26.23 & GW170814 \\\hline
Aug 17 2017 & 12:41:04 & $>1.86$ $10^{5}$ & 30.8 & 1.20 & GW170817 \\\hline
Aug 18 2017 & 02:25:09 & \multicolumn{3}{|c|}{found as a L single trigger but not as a coincidence} & GW170818 \\\hline
Aug 23 2017 & 13:13:58 & $14.7$ & 11.0 & 42.84 & GW170823 \\\hline
\end{tabular}
\end{table*}

\section{Conclusion and outlook} \label{conclusions}
The MBTA pipeline has been improved by including several new features, most notably the gating of suspected artifacts as part of data preprocessing, the extension of the search to higher masses, the definition of a single trigger ranking statistic blending in consistency with astrophysical signals as well as a pipeline-based witness of data quality, the inclusion of multi-detector parameters likelihood into the statistic used to rank coincidences, the consideration of heterogeneous detector sensitivities in sharing the background budget among coincidence types, the inclusion of sub-threshold information to improve sky localization. The pipeline was successfully used to perform a low-latency search of CBC signals in the LIGO and Virgo data during the O3 run, with sub-minute latency, and contributed to many alerts sent out by the LIGO-Virgo Collaboration.

For the first time in O3, the pipeline was further developed and adapted in order to perform offline searches as well.
This involves new features such as using future data in addition to past data to assess event significance, high-statistics injection runs, a dedicated system of scripts to handle the production, and a suite of post-processing macros to publish results on standardized web pages.
The effectiveness of the offline pipeline was illustrated on a representative subset of O2 data, with and without the injection of simulated signals.

With the exciting prospects of completing the analysis of O3 data and of starting the O4 run in 2022 with improved detectors, new challenges are lying ahead for the MBTA pipeline. To name a few prominent goals, let us mention estimating the probability for event candidates to be of astrophysical origin, extending the parameter space both to lower and higher masses, assessing the significance of single-detector triggers, including more features in signal waveforms, using a finer segmentation of the parameter space for false alarm estimation, reducing the latency of the online search and exploring the possibility of pre-merger alerts.  With an eye to the post-O4 era and the third generation of ground-based detectors, we note that MBTA is well suited to a detection bandwidth extended toward lower frequencies, through its multi-band feature which could be leveraged with more than the current two bands.

\section*{Acknowledgments}
We express our gratitude to Gijs Nelemans and John T. Whelan, who have dedicated time and energy to perform internal LIGO-Virgo review of the MBTA pipeline.
We thank our LIGO-Virgo collaborators from the CBC and low-latency groups for constructive comments.
This analysis exploits the resources of the computing facility at the EGO-Virgo site, and of the Computing Center of the Institut National de Physique Nucl\'e{}aire et Physique des Particules (CC-IN2P3/CNRS).
This research has made use of data, software and/or web tools obtained from the Gravitational Wave Open Science Center (https://www.gw-openscience.org), a service of LIGO Laboratory, the LIGO Scientific Collaboration and the Virgo Collaboration. LIGO Laboratory and Advanced LIGO are funded by the United States National Science Foundation (NSF) as well as the Science and Technology Facilities Council (STFC) of the United Kingdom, the Max-Planck-Society (MPS), and the State of Niedersachsen/Germany for support of the construction of Advanced LIGO and construction and operation of the GEO600 detector. Additional support for Advanced LIGO was provided by the Australian Research Council. Virgo is funded, through the European Gravitational Observatory (EGO), by the French Centre National de Recherche Scientifique (CNRS), the Italian Istituto Nazionale della Fisica Nucleare (INFN) and the Dutch Nikhef, with contributions by institutions from Belgium, Germany, Greece, Hungary, Ireland, Japan, Monaco, Poland, Portugal, Spain.

\appendix
\section{Pipeline online and offline implementation} \label{app:implementation}
\subsection{Online implementation}
The online pipeline is operated via the VPM custom tool, a graphical user interface common to the Virgo online system.
It allows process start/stop, live reporting of process status, editing and logging of the configuration files.
To provide full tracking, all VPM actions and key process activities are logged.
During the second half of O3, the full pipeline was running on five dedicated Linux machines at the computing facility at the EGO-Virgo site, using a total of 150 cores.
The typical CPU load was around 60\%.

\subsection{Offline implementation}
The offline implementation uses version 4 of the code for the full processing of O3 data. The main differences with
respect to the online concern the adaptation to massively run on data with the addition of simulated signals and several optimizations in
order to favour the detection efficiency over latency requirements. In the offline version of the code the update
of the PSD is done on a longer period than online and the FAR is estimated by using the
full data taking period of interest (typically 6 days).

The filtering step is run in parallel on the full preprocessed $h(t)$ for the standard search, and on a copy
of it with superimposed injections for performance studies.
The FAR vs $\rho_{RS}$ relation is determined on the run without injections. The relation is then used to associate a FAR value
to events in the data and the simulation runs.
In the offline setup the FAR is also monitored as a function of time in sub-periods of 6~hours to check the stability of
the noise and running conditions.
The analysis output is post-processed with a set of ROOT macros ~\cite{ROOT} to automatically produce relevant histograms and the list
of candidates with their characteristics. These results are then published on a set of web pages for further scrutiny.

The full code is run on 8-core jobs over the batch farm at the CC-IN2P3 computing center~\cite{CCIN2P3}. The runs are organised
by a semi-automatic set of bash scripts handling the automatic generation of configuration files, the
several-step submission procedure, the collection of the output and the bookkeeping. Special care is given to the job
planning to comply with the CPU, disk space and memory limits of the batch system, and to maximise
the CPU usage efficiency. The processing of one week of coincidence time in the three search regions can
be approximately quantified as 6~Ms, 26~Ms and 70~Ms of CPU time on one-core equivalent for BNS, BBH and NSBH, respectively.
These times include runs on data and injections. The template bank splitting for the filtering is organized in
about 200 jobs with a typical wall-clock time to completion between 15 and 40 hours, and a CPU usage
efficiency between 60\% and 100\%, depending on the job under consideration.

\end{document}